\documentclass{article}

\usepackage[preprint]{neurips_2026}
\setcitestyle{numbers,square,comma}
\usepackage[utf8]{inputenc} 
\usepackage[T1]{fontenc}    
\usepackage{hyperref}       

\usepackage{url}            
\usepackage{booktabs}       
\usepackage{amsmath}        
\usepackage{amsfonts}       
\usepackage{nicefrac}       
\usepackage{microtype}      
\usepackage{xcolor}         
\usepackage{graphicx}       
\usepackage{multirow}       
\usepackage{makecell}       
\usepackage{adjustbox}      
\usepackage{tikz}           
\usetikzlibrary{positioning}

\usepackage{booktabs}
\usepackage{tabularx}
\usepackage{wrapfig}
\usepackage{tcolorbox}
\usepackage{enumitem}
\usepackage{placeins}

\title{A Few Steps Further: Why Defenses Against Malicious Finetuning Erode Under Continued Training}

\author{\textbf{Itay Zloczower}$^{*}$, \textbf{Eyal Lenga}$^{*}$, \textbf{Gilad
 Gressel}$^{\dagger}$, \textbf{Yisroel Mirsky}$^{*}$ \\
 $^{*}$Ben-Gurion University of the Negev, $^{\dagger}$Amrita Vishwa Vidyapeetham, Amritapuri \\
 \{itayzloc, lenga\}@post.bgu.ac.il, gilad.gressel@am.amrita.edu,
 yisroel@bgu.ac.il
 }

\begin{document}

\maketitle

\begin{abstract}
Model providers increasingly release the weights of large language models. Although these models are safety-aligned before release, their safeguards can often be removed by fine-tuning on harmful data. A growing class of defenses aims to make alignment robust to such malicious fine-tuning, but these defenses are typically evaluated against attacks with a fixed training budget, even though an attacker who holds the weights can simply train for longer. We ask whether current defenses withstand this simplest escalation. Surveying fifteen recent defenses, we find that they share a common weakness: each is built around a limited model of the attacker, such as a bounded perturbation, a short simulated attack, or a trained link between harmful and benign behavior, and nothing enforces that protection once the weights are released. We then test six representative defenses on four open-weight models by continuing the same harmful-only fine-tuning attack for three epochs and measuring harmfulness and capability along the way. In all 72 defended runs, the model is more harmful at the end of training than at release, and on Llama-3.1 at the highest learning rate the defended models end almost as harmful as the undefended one. The defenses are not equally weak: one defense kept harmfulness low on one model, and some attacks recovered harmfulness only at the cost of general capability. Current defenses can delay or disrupt malicious fine-tuning, but in most cases their measured resistance does not persist under continued training, and they should not yet be treated as durable protection.
\end{abstract}

\section{Introduction}

Open-weight large language models (LLMs) create a post-release safety problem. Providers use methods such as preference alignment or reinforcement learning from human feedback (RLHF) to teach models safe behavior, and attackers can use their own data to try to undo that alignment. Prior work shows that small malicious fine-tuning (MFT) runs can subvert aligned models and recover harmful behavior \citep{yang2023shadow, lermen2023lora}; even benign fine-tuning can degrade safety alignment \citep{qi2023fine}. This fact reflects a structural asymmetry: the defender acts once, before release, while the attacker controls every update afterwards, including how long to keep training.

This risk has given rise to a growing body of MFT defenses: methods that aim to make harmful fine-tuning fail, require prohibitive optimization effort, or destroy the model's general capability when the attack succeeds. Their mechanisms vary widely. Some harden internal representations against harmful updates \citep{huang2024vaccine, rosati2024representation, liu2025targeted, chen2025vulnerability, perin2025lox}; some simulate adversarial fine-tuning during alignment \citep{henderson2023self, tamirisa2024tamper, sanyal2026antidote}; some reshape the loss landscape around safety-relevant parameters or optimization trajectories \citep{huang2024booster, rosati2025locking, fan2025towards, nguyen2026antibody}; and some deliberately couple harmful fine-tuning to utility collapse \citep{chen2025sdd, yi2025ctrap, wang2025self}. Despite this diversity, their evaluations share a common pattern: the attacker is a fixed supervised fine-tuning run with a prescribed loss, optimizer, data budget, and hyperparameter range \citep{qi2023fine, huang2024vaccine, rosati2024representation, tamirisa2024tamper, huang2024booster}.

This evaluation practice conflicts with a central lesson from adversarial machine learning: defenses must be evaluated against attackers who are free to deviate from the attack the defense was designed around, because robustness shown against one fixed attack routinely disappears once the attacker does something slightly different \citep{carlini2017towards, athalye_obfuscated_gradients, tramer_adaptive_attacks}. For MFT defenses, Qi et al.~\citep{qi2025evaluating} have already shown this for RepNoise and TAR: changing the random seed, trainer, schedule, or data overturns conclusions drawn from the original evaluations. We ask a simpler question: what happens when the attacker trains for longer? A natural response is that the defender can also train for longer. However, the two parties do not act symmetrically. The defender moves first and commits to a fixed defense before release; the attacker moves second, observes the released weights, and chooses a training budget in response. A durable defense therefore cannot rely on out-training the attacker. Its protection must hold at budgets beyond those used to construct and evaluate it.

\noindent\textbf{Many defenses, one shared limitation.}
A useful MFT attack must make the model harmful while preserving its general capability, so a defense can succeed by disrupting either condition. We find that fifteen recent MFT defenses fall into two broad strategies: \emph{anchoring}, which makes optimization on harmful objectives ineffective, and \emph{self-destruction}, which lets that optimization proceed but makes the model's capability collapse. Despite their surface differences, these defenses reduce to four loss templates. Each template protects a bounded perturbation region, a representation present at defense time, a short simulated attack, or a trained coupling between harmful and benign behavior. In every case, the protection is built around a finite model of the attacker, and nothing enforces it once the weights are released. With enough training, the model moves beyond the region or attack horizon the defense was built for, and the protection no longer applies.

\noindent\textbf{Continued training erodes measured resistance.}
We test this prediction by applying six defenses, spanning both strategies and three of the four templates, to four open-weight models: Llama-2-7B-chat~\citep{touvron2023llama}, Qwen3-8B~\citep{yang2025qwen3}, Llama-3.1-8B-Instruct~\citep{grattafiori2024llama}, and Zephyr-7B-$\beta$~\citep{tunstall2023zephyr}. We then run the same harmful-only fine-tuning attack on each defended model and on its undefended counterpart for three epochs at three learning rates, recording harmfulness and capability at 21 checkpoints. In all 72 defended trajectories, harmfulness at the final checkpoint exceeds harmfulness at release, with a mean increase of $0.372$ in StrongREJECT score~\citep{souly2024strongreject}. On Llama-3.1-8B at the highest learning rate, the six defended models finish within $0.013$ of the harmfulness reached by the undefended model under the same attack. The effect is not uniform: VAA~\citep{chen2025vulnerability} keeps harmfulness low on Zephyr-7B-$\beta$ at every checkpoint, and in several trajectories harmfulness rises only alongside a loss of capability. In most trajectories, however, the defenses delay harmful recovery rather than prevent it.

\noindent\textbf{This work makes two contributions.}
\textbf{(1)}~We systematize fifteen recent MFT defenses (twelve since 2025) and show that, despite their apparent diversity, they reduce to two strategies and four loss templates, each of which protects the model only within the region, representation, or attack horizon it encodes. \textbf{(2)}~We show empirically that continued fine-tuning under a fixed, harmful-only objective erodes measured resistance for every evaluated defense, across 72 defended trajectories on four models. Together, these results indicate that resistance to a fixed fine-tuning budget should not be interpreted as durable protection for open-weight models.

\section{Related Work}

\noindent\textbf{Durable safeguards for open-weight models.}
A growing line of work studies whether safety interventions remain effective once a model is released as open weights and exposed to downstream fine-tuning. Standard refusal training can be weakened or removed by continued optimization \citep{qi2023fine, yang2023shadow, zhan_removing_safety, wei_pruning_safety}, which has motivated defenses that aim to make harmful behavior difficult to recover after release. These include representation noising, tamper-resistant meta-learning, adversarial unlearning, and other training-time interventions intended to make MFT fail or degrade utility \citep{rosati2024representation, tamirisa2024tamper, li_wmdp, huang2024vaccine}. \autoref{sec:systematization} organizes fifteen such defenses into a common taxonomy.

\noindent\textbf{Evaluating durability.}
The closest prior work is Qi et al.~\citep{qi2025evaluating}, who show that MFT defenses can be far more brittle than their original evaluations suggest. For RepNoise and TAR, repeating the attack across random seeds, or changing the trainer implementation, learning-rate schedule, prompt template, or fine-tuning data, overturns conclusions drawn from the original evaluations. Their lesson is that a defense should be stress-tested across plausible downstream configurations before its weights are released. Our work asks what remains once those problems are addressed. Suppose a defense has been implemented correctly and is effective against the attack it is evaluated with. It still has to hold when the attacker trains for longer, because the training budget, unlike the defense, is chosen after release. This is the question we test. We first construct each defense, choosing a defense configuration that reduces harmfulness while retaining capability, and then continue the same attack, with the configuration held fixed, for three full epochs, measuring whether the defense's effect persists. We also extend the evaluation from two defenses to six, drawn from a taxonomy of fifteen.

\noindent\textbf{Attack budgets in adversarial machine learning.}
A central lesson from adversarial machine learning is that robustness claims based on static or otherwise non-adaptive attacks often fail under adversaries that optimize against the defense itself \citep{biggio_evasion_2013, szegedy_intriguing_2014, goodfellow_explaining_2015, carlini_wagner_detection, athalye_obfuscated_gradients, tramer_adaptive_attacks, mujkanovic_gnn_robustness}. One of the simplest forms of this failure is an under-powered attack. Carlini et al.~\citep{carlini2019evaluating} recommend plotting attack success against the number of attack iterations and verifying that doubling the iterations does not increase success, observing that there are few reasonable threat models in which an attacker can afford 100 iterations but not 1{,}000. The analogous quantity for MFT is the fine-tuning budget, and it is controlled entirely by the attacker. Our cost curves apply this check to MFT defenses: we track harmfulness and capability as the number of attacker updates grows, rather than reporting a single stopping point.

\noindent\textbf{Adaptive attacks in LLM security.}
Similar evaluation failures have recently appeared in LLM security. Jailbreak and prompt-injection defenses are often evaluated against fixed benchmark attacks, static prompt sets, or generic optimization methods that are not tuned to the defense, and can appear robust under these evaluations while failing against attackers that adapt their search procedure, feedback signal, or strategy \citep{nasr_attacker_moves_second, zou_gcg, chao_pair, mazeika2024harmbenchstandardizedevaluationframework, debenedetti_agentdojo}. Our setting differs in the object being attacked: rather than constructing adversarial inputs for a deployed model, the adversary modifies the weights directly. Such an attacker can choose the data mixture, objective, optimizer, schedule, and training duration. We deliberately vary only the last of these, using a common harmful-only objective and a fixed dataset, so that any erosion we observe cannot be attributed to a stronger or defense-specific attack.

\section{Systematizing MFT Defenses}
\label{sec:systematization}
\subsection{Problem Setup}
\label{sec:problem-setup}

\noindent\textbf{Threat model.} A defender releases the weights $\theta_{\mathrm{def}}$ of an aligned model. The attacker has white-box access to these weights and full control over subsequent fine-tuning, including the data, objective, optimizer, learning-rate schedule, and number of updates. The attacker's goal is a model that complies with harmful requests while retaining its general capability; a model that produces harmful content but has lost fluency, coherence, or usefulness does not meet this goal. A defense therefore succeeds if it prevents either condition, and the two strategies below correspond to these two options. Some methods were developed for fine-tuning services, where the provider can also control later training. Here we consider the components that can be applied before releasing the weights.

\noindent\textbf{Loss definitions.} For language models, harmful loss $\mathcal{L}_h$ is the model's average prediction loss on harmful prompt--response pairs, where the target response fulfills the harmful request. Capability loss $\mathcal{L}_c$ is its average prediction loss on tasks that measure general capability. Lower harmful loss means the model fits harmful responses more closely. Lower capability loss means better performance on the capability tasks.

\subsection{Fifteen Defenses, Two Strategies}
\begin{figure}[!htbp]
\centering
\includegraphics[width=\linewidth]{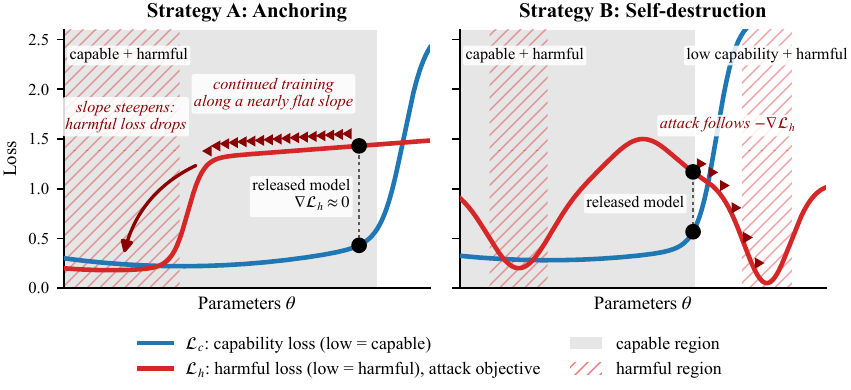}
\caption{Schematic examples of the two strategies. \textbf{Anchoring (left):} the released model lies on a nearly flat stretch of harmful loss $\mathcal{L}_h$. The illustrated attack initially makes little progress, but continued training moves it toward a steeper region where harmful loss falls. This is one possible mechanism, not a property established for every anchoring method. \textbf{Self-destruction (right):} harmful loss falls while capability loss $\mathcal{L}_c$ rises, leaving the attacker with a less capable model.}
\label{fig:defense-strategies}
\end{figure}

A successful attack needs two things: the fine-tuned model must be harmful, and it must remain capable. MFT defenses try to deny one of the two. \emph{Anchoring} defenses aim to prevent the model from learning harmful behavior. \emph{Self-destruction} defenses aim to make harmful fine-tuning damage general capability, so the attacker cannot obtain a model that is both harmful and useful. Individual methods can pursue both aims. Our taxonomy groups fifteen methods by their main intended effect, including earlier work on blocking harmful tasks. \autoref{fig:defense-strategies} illustrates the two strategies, and \autoref{tab:mft-defense-taxonomy} gives the assignments.

\begin{table}[!t]
\centering
\small
\setlength{\tabcolsep}{6pt}
\renewcommand{\arraystretch}{.9}
\begin{tabular}{@{}llcc@{}}
\toprule
Sub-strategy & Defense & Year & Template \\
\midrule
\multicolumn{4}{@{}l}{\textit{\textbf{Strategy A: Anchoring}}} \\
\midrule
Representation & Vaccine \cite{huang2024vaccine} & 2024 & T1 \\
               & T-Vaccine \cite{liu2025targeted} & 2025 & T1 \\
               & RepNoise \cite{rosati2024representation} & 2024 & T2 \\
\addlinespace[2pt]
Direction      & VAA \cite{chen2025vulnerability} & 2025 & T1 \\
               & SAM-unlearning \cite{fan2025towards} & 2025 & T1 \\
               & Antibody \cite{nguyen2026antibody} & 2026 & T3 \\
               & Booster \cite{huang2024booster} & 2025 & T3 \\
\addlinespace[2pt]
Subspace       & LoX \cite{perin2025lox} & 2025 & T1 \\
               & AntiDote \cite{sanyal2026antidote} & 2026 & T1 \\
\addlinespace[2pt]
Objective      & KT-IPA \cite{cheng2025weaponization} & 2025 & T2\,+\,T3 \\
\addlinespace[2pt]
Trajectory     & MLAC \cite{henderson2023self} & 2023 & T3 \\
               & TAR \cite{tamirisa2024tamper} & 2025 & T3 \\
\midrule
\multicolumn{4}{@{}l}{\textit{\textbf{Strategy B: Self-destruction}}} \\
\midrule
---            & CTRAP \cite{yi2025ctrap} & 2026 & T3 \\
               & SEAM \cite{wang2025self} & 2026 & T4 \\
               & SDD \cite{chen2025sdd} & 2025 & T4 \\
\bottomrule
\end{tabular}
\caption{Taxonomy of fifteen methods. Strategies describe the intended effect on an attack; T1--T4 denote loss templates; KT-IPA combines two. Antibody's placement concerns its alignment stage. LoX is a training-free analogue of T1.}
\label{tab:mft-defense-taxonomy}
\end{table}

\noindent\textbf{Strategy A: anchoring.} Anchoring defenses aim to keep harmful loss $\mathcal{L}_h$ high when an attacker fine-tunes the model to lower it. Some methods seek to weaken the harmful gradient, which can slow gradient-based learning. The plateau in \autoref{fig:defense-strategies} illustrates this possible mechanism. Other methods train refusals under weight or activation changes, or remove information used for harmful tasks. These methods need not produce the same loss landscape; for example, TAR's refusal objective does not necessarily flatten harmful loss \citep{tamirisa2024tamper}.

\emph{Representation anchoring} acts on internal activations. Vaccine and T-Vaccine train on refusals while perturbing activations, aiming to preserve refusals after harmful fine-tuning \citep{huang2024vaccine,liu2025targeted}. RepNoise pushes harmful-response activations toward random noise to make harmful behavior harder to recover \citep{rosati2024representation}. \emph{Direction anchoring} changes how the model responds to training updates. Booster penalizes the decrease in harmful loss after a simulated harmful step \citep{huang2024booster}. Antibody's alignment stage uses a related penalty and trains the updated model to produce a generic refusal. Its full method also reweights examples during downstream fine-tuning, a stage that requires provider control and falls outside our open-weight setting \citep{nguyen2026antibody}. SAM-unlearning trains its unlearning objective under bounded weight changes \citep{fan2025towards}. VAA applies weight perturbations while giving more attention to the group of alignment examples on which the model currently performs worse \citep{chen2025vulnerability}.

Among the remaining methods, LoX strengthens the main directions in the weight change produced by safety alignment, and AntiDote trains against generated adversarial low-rank adaptation (LoRA) patches \citep{perin2025lox,sanyal2026antidote}. KT-IPA combines changes to internal activations with adversarial training \citep{cheng2025weaponization}. MLAC and TAR train against simulated attacks, aiming to preserve resistance after the attacker updates the model \citep{henderson2023self,tamirisa2024tamper}.

\noindent\textbf{Strategy B: self-destruction.} These defenses aim to make harmful fine-tuning increase capability loss $\mathcal{L}_c$. The attacker may then fail to obtain a useful model even if harmful loss decreases (\autoref{fig:defense-strategies}, right). CTRAP simulates one harmful fine-tuning step during alignment and trains the resulting model to output a fixed error token on general dialogue data, aiming to make later harmful updates damage capability \citep{yi2025ctrap}. SEAM trains the harmful and capability gradients to oppose each other. It also trains refusals and increases harmful loss, so resistance and self-destruction can be combined \citep{wang2025self}. SDD trains the model to answer harmful prompts with fluent but unrelated benign responses, so that fine-tuning it toward harmful answers is intended to damage general instruction following \citep{chen2025sdd}. Capability loss must therefore be evaluated during the attack; a decrease in harmful loss alone does not establish that the attacker has succeeded.

\subsection{Four Loss Templates Behind the Taxonomy}
\label{sec:templates}
The anchoring/self-destruction axis describes the \emph{intended failure mode} for the attacker. A second axis describes \emph{how} each defense encodes the future attacker in its training objective. The fifteen defenses fall into four templates. Moreover, every template builds on a standard alignment loss $\mathcal{L}_{\mathrm{align}}$, whose purpose is to keep the released model both safe and useful. It decomposes as $\mathcal{L}_{\mathrm{align}}(\theta) = \mathcal{L}_s(\theta) + \mathcal{L}_c(\theta)$: a safety term $\mathcal{L}_s$ (refusal on harmful prompts) and a capability term $\mathcal{L}_c$ (correct behavior on benign prompts). A defender that dropped $\mathcal{L}_c$ would release a useless model, so every defense in our survey includes it. \noindent\textbf{Template 1: Robust-alignment basin.} A first family tries to place the released model in a basin where alignment is stable to bounded perturbations:
\[
\mathcal{L}_{\mathrm{def}}^{(1)}(\theta)
=
\max_{\delta\in\Delta}
\mathcal{L}_{\mathrm{align}}(\theta+\delta)
+
\lambda \mathcal{L}_{c}(\theta).
\tag{T1}
\]
The perturbation set $\Delta$ may live in embedding space, layer space, or weight space. Vaccine uses adversarial hidden-state perturbations; T-Vaccine

restricts them to layers selected by harmful-gradient norm; VAA combines group-level robustness with adversarial sampling; SAM-unlearning uses a sharpness-aware neighborhood; and Antibody's alignment-stage flatness term is a related local robustness objective \cite{huang2024vaccine,liu2025targeted,chen2025vulnerability,fan2025towards,nguyen2026antibody}. LoX is the post-hoc analogue: instead of solving (T1), it extrapolates along an estimated low-rank safety direction to move the model into a flatter region \cite{perin2025lox}. Template 1 implements anchoring. It does not remove harmful capability globally; it makes the \textit{local} basin around $\theta_{\mathrm{def}}$ harder to exit by ordinary harmful fine-tuning.

\noindent\textbf{Template 2: Harmful-information removal.} A second family attacks the information content of harmful representations:
\[
\mathcal{L}_{\mathrm{def}}^{(2)}(\theta)
=
\mathcal{L}_{\mathrm{align}}(\theta)
-
\beta \mathcal{L}_h(\theta)
+
\alpha R_{\mathrm{purge}}(\theta;\mathcal{D}_h).
\tag{T2}
\]
Here $R_{\mathrm{purge}}$ removes useful structure from activations on harmful inputs. RepNoise instantiates this with a noise-matching regularizer,

using MMD to push harmful activations toward random noise \cite{rosati2024representation}. KT-IPA includes a related representation purging stage before applying its prospect-theoretic integrity objective \cite{cheng2025weaponization}. Template 2 also implements anchoring. Rather than flattening a basin, it tries to make the internal features that $\nabla \mathcal{L}_h$ would manipulate less useful, so the harmful gradient at $\theta_{\mathrm{def}}$ carries no usable signal.
\noindent\textbf{Template 3: Look-ahead defense.} 
The most general template explicitly simulates a future attacker:
\[
\mathcal{L}_{\mathrm{def}}^{(3)}(\theta)
=
\mathcal{L}_{\mathrm{align}}(\theta)
+
\lambda R(\theta'),
\qquad
\theta'=\texttt{SimAttack}(\theta;\mathcal{D}_h).
\tag{T3}
\]
The simulated attack may be one harmful gradient step, $K$ inner fine-tuning steps, a sampled adversary, or a learned patch. The penalizer $R$ then imposes a desired property at the simulated post-attack point.

Template 3 is the bridge between anchoring and self-destruction. If $R$ penalizes harmful progress, preserves refusal, prefers safe responses, or keeps harmful loss high along a simulated trajectory, the method is an anchoring defense. Booster penalizes harmful-loss drop after a simulated harmful step \cite{huang2024booster}; Antibody preserves refusal at the post-step model \cite{nguyen2026antibody}; AntiDote trains against activation-conditioned adversarial LoRA patches \cite{sanyal2026antidote}; KT-IPA includes an adversarial phase inside its integrity objective \cite{cheng2025weaponization}; and MLAC and TAR penalize a $K$-step simulated trajectory so harmful loss stays high or the predictive distribution stays at maximum entropy \cite{henderson2023self,tamirisa2024tamper}. If $R$ instead makes the simulated endpoint useless on benign data, the same template implements self-destruction. CTRAP penalizes a simulated post-harmful-step model to collapse on benign inputs \cite{yi2025ctrap}. Thus the same mathematical form can implement either strategy. The difference is what the defender wants to be true at $\theta'$: a model that still refuses (anchoring), or a model that has lost benign capability (self-destruction).

\noindent\textbf{Template 4: Coupling trap.} The last template does not simulate the attacker. Instead, it directly couples harmful improvement to benign degradation:
\[
\mathcal{L}_{\mathrm{def}}^{(4)}(\theta)
=
\mathcal{L}_{\mathrm{align}}(\theta)
+
\mathcal{L}_{\mathrm{couple}}(\theta;\mathcal{D}_h,\mathcal{D}_c).
\tag{T4}
\]
where $\mathcal{D}_c$ is typically benign task samples. SEAM implements this explicitly by shaping the relationship between harmful

and benign gradients: descent on harmful data is made to increase benign loss \cite{wang2025self}. SDD implements a data-level version: it trains harmful prompts to elicit fluent but irrelevant benign answers, so later harmful fine-tuning must undo a response mapping that also damages instruction-following \cite{chen2025sdd}. Template 4 is always self-destruction. It does not try to hide the harmful gradient; it tries to make following it costly.

\FloatBarrier

\section{Experimental Setup}
\label{sec:setup}

\noindent\textbf{Question.} The argument in \autoref{sec:templates} predicts that a defense which resists MFT early in an attack will resist less as the attacker keeps training. We test this directly. We fix the attack data, objective, and hyperparameters, run each attack for three epochs, and measure harmfulness and capability at checkpoints along the way. We call the resulting curves \emph{cost curves}. Every defended model is compared with its undefended counterpart under the identical attack.

\noindent\textbf{Models and defenses.} We use four open-weight chat models: Llama-2-7B-chat~\cite{touvron2023llama}, Qwen3-8B~\cite{yang2025qwen3}, Llama-3.1-8B-Instruct~\cite{grattafiori2024llama}, and Zephyr-7B-$\beta$~\cite{tunstall2023zephyr}. On each, we build six defenses that cover both strategies and three of the four templates: Vaccine, VAA, and SAM-unlearning (T1); Booster and CTRAP (T3); and SDD (T4) \citep{huang2024vaccine,chen2025vulnerability,fan2025towards,huang2024booster,yi2025ctrap,chen2025sdd}. We use the authors' code where it is available. SAM-unlearning, labeled Unlearn-Smooth in the figures, was designed to unlearn WMDP-Bio knowledge; we port its NPO+SAM objective to all four models and set its forget target to harmful BeaverTails responses. For Vaccine, VAA, SDD, and SAM-unlearning, we select the defense configuration with a validation objective that combines harmfulness and capability. Booster and CTRAP use the final checkpoint of their training runs. \autoref{app:defense-provenance} gives the details.

\noindent\textbf{Attack.} The attacker fine-tunes the released model with LoRA on 9{,}000 harmful prompt--response pairs from BeaverTails~\cite{ji2023beavertails}, minimizing cross-entropy on the harmful responses alone. Each run lasts three epochs, or 3{,}375 optimizer steps at an effective batch size of eight, under a cosine learning-rate schedule with 337 warmup steps. We repeat every run at three peak learning rates, $10^{-5}$, $3\times10^{-5}$, and $10^{-4}$, and save 21 checkpoints spaced roughly logarithmically from step 1 to step 3{,}375. With six defended models and one undefended model per backbone, this gives 84 trajectories, 72 of them defended. We also attack four further undefended models (\autoref{app:undefended-curves}). \autoref{app:hyperparams} lists all hyperparameters.

\noindent\textbf{Metrics.} Harmfulness is the mean score of the fine-tuned StrongREJECT judge~\cite{souly2024strongreject} on a category-stratified sample of 1{,}000 held-out harmful BeaverTails prompts. The judge scores each response from 0 to 1 by how useful it is for the harmful request. Responses are decoded greedily with up to 512 new tokens. Capability is the zero-shot accuracy on validation subsets of MMLU~\cite{hendrycks2021mmlu}, TruthfulQA~\cite{lin2022truthfulqa}, HellaSwag~\cite{zellers2019hellaswag}, and ARC-Easy~\cite{clark2018arc}, 600 questions in total, averaged with equal weight per task (\autoref{app:setup_details}).

\section{Results}
\label{sec:cost-results}

\begin{figure}[tp]
\centering
\includegraphics[width=\linewidth]{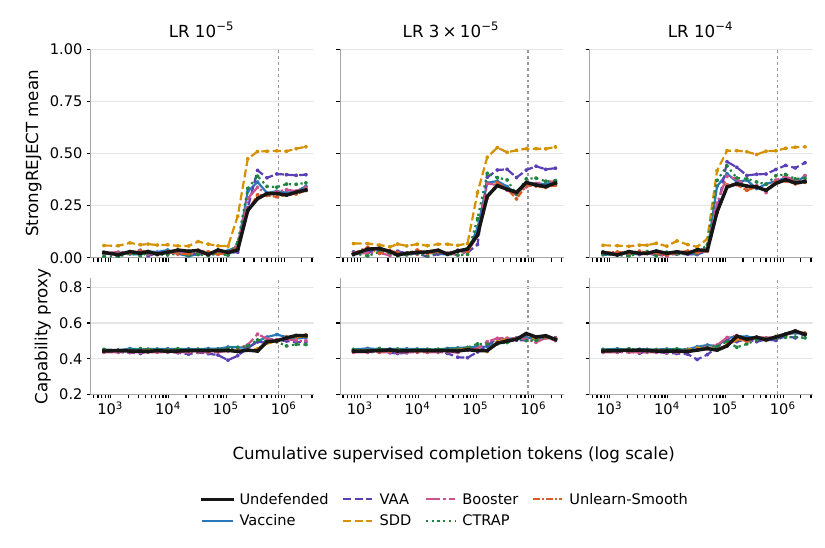}
\caption{Cost curves for Llama-2-7B-chat. Top row: harmfulness (mean StrongREJECT score); bottom row: capability. Columns correspond to peak learning rates. Each line joins the 21 checkpoints of one trajectory; black is the undefended model. The horizontal axis counts the supervised completion tokens the attacker has trained on, including repeats in later epochs, and the dotted vertical line marks the end of the undefended run's first epoch. The first checkpoint lies in the warmup at a learning rate of zero, so it shows the released model. Axis ranges are shared across Figures~\ref{fig:cost-llama2}--\ref{fig:cost-zephyr}.}
\label{fig:cost-llama2}
\end{figure}

\begin{figure}[tp]
\centering
\includegraphics[width=\linewidth]{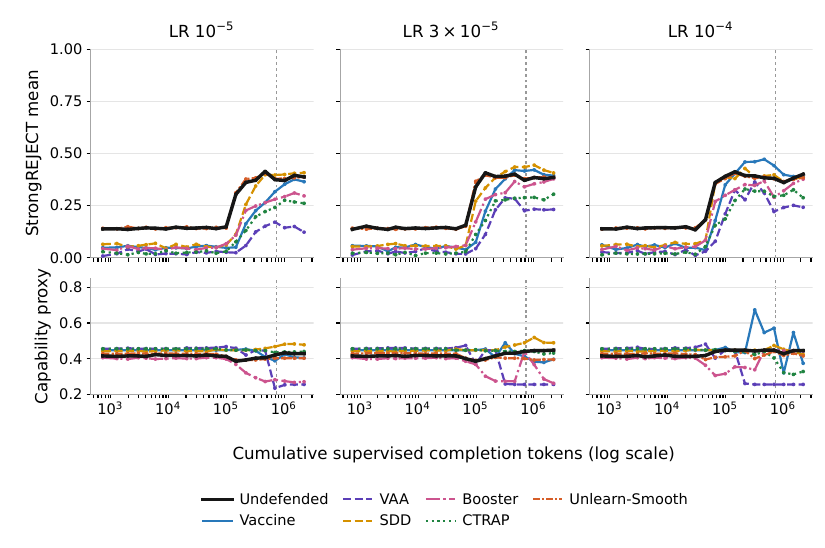}
\caption{Cost curves for Qwen3-8B, in the layout of \autoref{fig:cost-llama2}. VAA loses capability at every learning rate as harmfulness rises.}
\label{fig:cost-qwen3}
\end{figure}

\begin{figure}[tp]
\centering
\includegraphics[width=\linewidth]{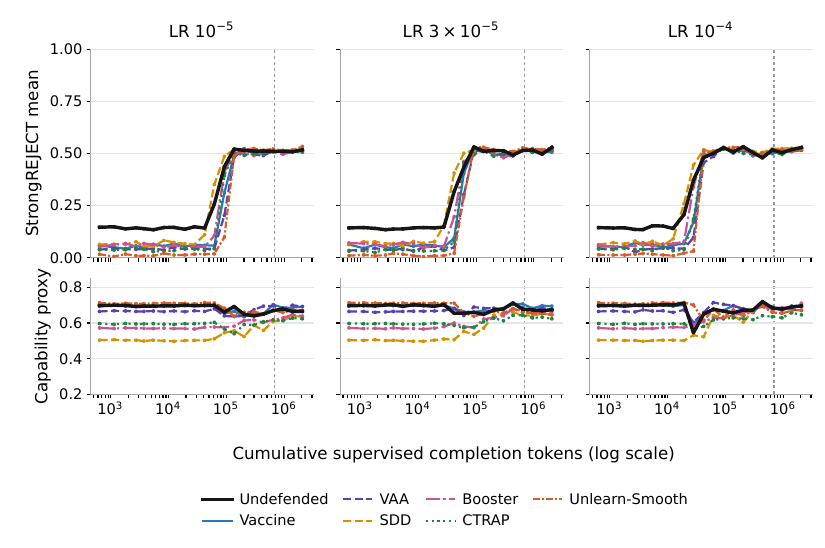}
\caption{Cost curves for Llama-3.1-8B-Instruct, in the layout of \autoref{fig:cost-llama2}. All defended models converge to the harmfulness of the undefended model.}
\label{fig:cost-llama31}
\end{figure}

\begin{figure}[tp]
\centering
\includegraphics[width=\linewidth]{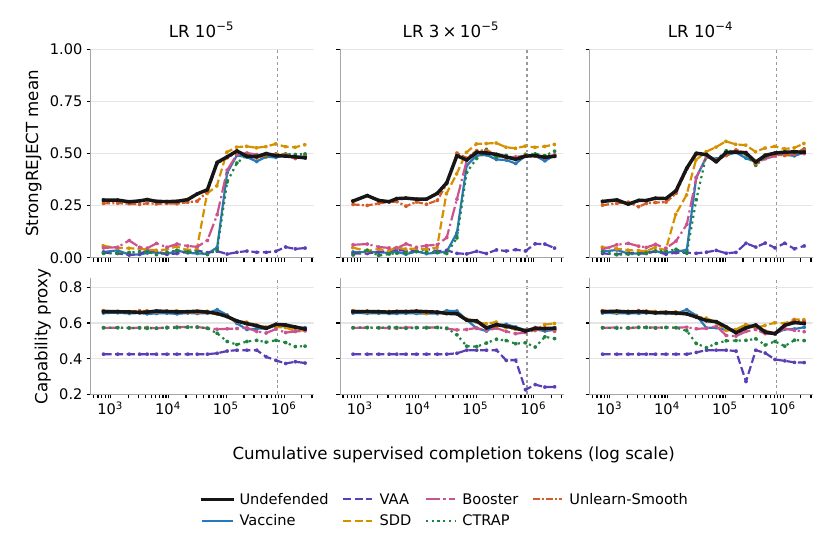}
\caption{Cost curves for Zephyr-7B-$\beta$, in the layout of \autoref{fig:cost-llama2}. VAA keeps harmfulness low throughout, at lower capability. The undefended and Unlearn-Smooth models are already more harmful at release than the other defended models.}
\label{fig:cost-zephyr}
\end{figure}

\noindent\textbf{Resistance erodes in every defended trajectory.} Figures~\ref{fig:cost-llama2}--\ref{fig:cost-zephyr} show the cost curves. In all 72 defended trajectories, harmfulness at the final checkpoint is higher than at release. The mean increase in StrongREJECT score is $0.372$, and individual increases range from $0.022$ to $0.520$. The pattern holds for all four models and for each evaluated template. In most defended trajectories, harmfulness rises within the first epoch, over the same range of training tokens as for the undefended model.

How far harmfulness rises depends on the model, defense, and learning rate. At the highest learning rate, Vaccine on Llama-2 starts at a harmfulness of $0.013$ and ends at $0.393$, while its capability rises from $0.451$ to $0.537$. On Llama-3.1 at the same learning rate, the six defended models end with harmfulness between $0.515$ and $0.527$, against $0.528$ for the undefended model. On this model, after three epochs, the defenses make almost no difference to how harmful the attacked model is. The exception is VAA on Zephyr, which never exceeds a harmfulness of $0.069$ at any learning rate. It starts with lower capability than the undefended Zephyr ($0.425$ against $0.664$) and ends at $0.378$ at the highest learning rate.

\noindent\textbf{Capability must be read alongside harmfulness.} Some attacks recover harmful behavior while capability falls. For CTRAP on Zephyr at the highest learning rate, harmfulness rises from $0.028$ to $0.521$ while capability falls from $0.574$ to $0.501$. For VAA on Qwen3 at the same learning rate, harmfulness peaks at $0.363$ and ends at $0.241$, while capability falls from $0.455$ to $0.255$. Harmfulness need not rise monotonically, and a harmful endpoint is not always a useful model for the attacker.

\section{Discussion}
\label{sec:discussion}

\noindent\textbf{What unifies the results.} The six defenses use different mechanisms, yet they erode in the way \autoref{sec:templates} predicts. Each protects the model near its released weights or along a short simulated attack, and continued training moves the model out of that region. The defenses differ in how the erosion shows: some delay the rise in harmfulness, some turn it into a loss of capability, and some end almost exactly where the undefended model does.

\noindent\textbf{Limitations.} Each trajectory uses a single seed, and the three learning rates are separate attack settings rather than replicates. The attack uses one dataset and one objective, and we vary how long the attacker trains, not how many distinct examples it sees. Our longest attack, three epochs over 9{,}000 examples, is itself a finite budget, so we cannot say whether a defense that holds within it, such as VAA on Zephyr, would hold under a longer attack. The capability measure is a 600-question proxy that was also used to select the defense configurations; it is not a substitute for full benchmark evaluation. We evaluate six of the fifteen defenses and none from T2. Finally, early checkpoints lie inside the warmup of a schedule set for the full run, so they do not show the least training an attacker with a shorter schedule would need.

\begin{tcolorbox}[
  colback=blue!5!white,
  colframe=blue!50!black,
  title=\textbf{Practical takeaway},
  fonttitle=\bfseries,
  boxrule=0.5pt,
  arc=2pt,
  left=6pt, right=6pt, top=4pt, bottom=4pt
]
Resistance to malicious fine-tuning at one training budget is not evidence of resistance at a larger one. The attacker chooses the budget after release, so a defense should be evaluated as the attack continues, with capability measured alongside harmfulness and an undefended model attacked with the same recipe as a reference.
\end{tcolorbox}

\section{Conclusion}
\label{sec:conclusion}

We systematized fifteen defenses against malicious fine-tuning into two strategies, anchoring and self-destruction, and four loss templates. Each template protects the model against a finite model of the attacker: a bounded perturbation, the representations present at defense time, a short simulated attack, or a trained coupling between harmful and benign behavior. None of these protections is enforced after release. Testing six defenses on four models, we found that continuing the same harmful fine-tuning attack raised harmfulness in all 72 defended trajectories; on Llama-3.1 at the highest learning rate, every defended model ended within $0.013$ of the undefended model. In most cases, current defenses delay malicious fine-tuning rather than prevent it.

\enlargethispage{2\baselineskip}
\section*{Acknowledgments}
This work was funded by the European Union through ERC grant AGI-Safety (101222135). Views and opinions expressed are, however, those of the authors only and do not necessarily reflect those of the European Union or the European Research Council Executive Agency. Neither the European Union nor the granting authority can be held responsible for them.

\bibliographystyle{plain}   
\bibliography{references}   

\newpage
\appendix

\section{Ethics and dual-use considerations}
\label{app:ethics}
This work studies how defenses against MFT behave as an attacker continues training, and is therefore dual-use: the same analysis that improves evaluation could also inform attackers. The attack uses public data and standard fine-tuning operations available to a holder of open weights. Our goal is to test whether conclusions about resistance persist beyond a short training budget. We report aggregate harmfulness and capability measurements rather than harmful outputs or deployment guidance.

\paragraph{Safeguards.}
All attack experiments use existing harmful BeaverTails examples; we contribute no new harmful corpus. Defense construction also uses public data, including Alpaca, and generated refusal responses where required by the defense protocol. We do not release defended or attacked model checkpoints. Code release covers the training and evaluation pipeline, including the defense adaptations needed to reproduce the evaluated checkpoints.

\section{LLM usage}
\label{app:llm-usage}
We use GPT-4o-mini to generate refusal responses for defense construction on a paired BeaverTails split disjoint from the attack data. This data-preparation step is separate from the threat model: the attacker neither queries nor fine-tunes a hosted model, and these refusals are not used in the attack. We use the locally run, fine-tuned StrongREJECT judge~\citep{souly2024strongreject} to score saved responses to held-out BeaverTails prompts, applying the same continuous scoring procedure to every trajectory. We also use LLMs for writing assistance, checking descriptions against cited sources, and drafting the paper's explanatory argument.
\section{Licenses for existing assets}
\label{app:licenses}
We list the licenses declared for the models, datasets, and defense code used in this work. All experiments are for non-commercial academic research.

\paragraph{Models.}
\begin{itemize}
\item \textbf{Llama-2-7B-chat and Llama-2-13B-chat}~\cite{touvron2023llama}: Meta Llama 2 Community License Agreement.
\item \textbf{Llama-3.1-8B-Instruct}~\cite{grattafiori2024llama}: Meta Llama 3.1 Community License Agreement.
\item \textbf{Llama-3.2-3B-Instruct}: \href{https://huggingface.co/meta-llama/Llama-3.2-3B-Instruct}{Llama 3.2 Community License Agreement}.
\item \href{https://huggingface.co/Qwen/Qwen3-8B}{\textbf{Qwen3-8B}}~\cite{yang2025qwen3}, \href{https://huggingface.co/Qwen/Qwen2.5-7B-Instruct}{\textbf{Qwen2.5-7B-Instruct}}, and \href{https://huggingface.co/mistralai/Mistral-7B-Instruct-v0.3}{\textbf{Mistral-7B-Instruct-v0.3}}: Apache License 2.0.
\item \textbf{Zephyr-7B-$\beta$}~\cite{tunstall2023zephyr}: \href{https://huggingface.co/HuggingFaceH4/zephyr-7b-beta}{MIT License}.
\item \textbf{StrongREJECT judge}~\cite{souly2024strongreject}: the \href{https://huggingface.co/qylu4156/strongreject-15k-v1}{adapter model card} does not specify a license. Its Gemma-2B base is subject to the \href{https://ai.google.dev/gemma/terms}{Gemma Terms of Use}.
\end{itemize}

\paragraph{Datasets.}
\begin{itemize}
\item \textbf{BeaverTails}~\cite{ji2023beavertails}: CC BY-NC 4.0. We use only the publicly released splits and use the data exclusively for non-commercial research.
\item \textbf{Alpaca}~\cite{taori2023alpaca}: CC BY-NC 4.0. Use is restricted to non-commercial research.
\item \textbf{MMLU}~\cite{hendrycks2021mmlu} and \textbf{HellaSwag}~\cite{zellers2019hellaswag}: MIT License.
\item \textbf{TruthfulQA}~\cite{lin2022truthfulqa}: Apache License 2.0.
\item \textbf{ARC}~\cite{clark2018arc}: CC BY-SA 4.0.
\end{itemize}

\paragraph{Defense code.}
We use author-released defense code and wrappers, with the NPO+SAM adaptation described in \autoref{app:defense-provenance}. Vaccine~\cite{huang2024vaccine}, Booster~\cite{huang2024booster}, and SDD~\cite{chen2025sdd} are released under Apache License 2.0. SAM-unlearning (Unlearn-Smooth)~\cite{fan2025towards} is released under the MIT License. VAA~\cite{chen2025vulnerability} and CTRAP~\cite{yi2025ctrap} do not have a license file declared in their public repositories; we use them solely for non-commercial academic research and replication, and we cite the original papers. Data adapters convert the shared data into the required formats; our SAM-unlearning port also changes the trainer, backbone support, and unlearning task.

\paragraph{Defense-data preparation service.}
We use OpenAI's GPT-4o-mini API only to generate refusal responses for defense construction on the paired BeaverTails split (see \autoref{app:llm-usage}). This is not an attack interface; all attack fine-tuning, response generation, and harmfulness judging run locally.

\section{Extended Defense Taxonomy}
\label{app:taxonomy}
\autoref{tab:mft-defense-taxonomy-appendix} records the author-intent rationale behind each placement in the main taxonomy. It summarizes the property each method tries to install.

\begin{table}[p]
\centering
\footnotesize
\setlength{\tabcolsep}{3pt}
\renewcommand{\arraystretch}{1.25}
\begin{tabular}{@{}p{1.9cm}p{2.2cm}p{1.1cm}p{7.3cm}@{}}
\toprule
Sub-strategy & Defense & Template & Author intent \\
\midrule
\multicolumn{4}{@{}l}{\textit{\textbf{Strategy A: Anchoring}}} \\
\midrule
Representation
& Vaccine \cite{huang2024vaccine}
& T1
& Add bounded adversarial perturbations to hidden embeddings at every layer during alignment, so the aligned solution sits in a basin that resists the embedding drift caused by MFT. \\
& T-Vaccine \cite{liu2025targeted}
& T1
& Use the same robust-basin idea as Vaccine, but perturb and train only a subset of layers, sampled in proportion to their harmful-data gradient norms. \\
& RepNoise \cite{rosati2024representation}
& T2
& Push hidden activations on harmful inputs toward Gaussian noise using a per-layer MMD term, removing structure that a harmful gradient could manipulate. \\
\addlinespace[3pt]
Direction
& VAA \cite{chen2025vulnerability}
& T1
& Split the alignment data into examples that are and are not forgotten during a proxy harmful fine-tune, then apply group DRO with SAM-style weight perturbations to upweight the vulnerable group. \\
& SAM-unlearning \cite{fan2025towards}
& T1
& Apply sharpness-aware minimization to the forget loss, so the unlearned solution lies in a flat region that resists relearning. \\
& Antibody \cite{nguyen2026antibody}
& T3
& Flatten the harmful loss around the defended weights and preserve the likelihood of a generic refusal, both evaluated after one simulated harmful gradient step. \\
& Booster \cite{huang2024booster}
& T3
& Penalize the drop in harmful loss after one normalized harmful gradient step. To first order, this penalizes the norm of the harmful gradient. \\
\addlinespace[3pt]
Subspace
& LoX \cite{perin2025lox}
& T1$^{\dagger}$
& Amplify the top singular subspace of the safety-alignment weight update, moving the released weights toward a flatter region of the safety landscape without further training. \\
& AntiDote \cite{sanyal2026antidote}
& T1
& Train against an activation-conditioned hypernetwork that generates adversarial LoRA patches, requiring the model to keep its safety preference under the patches while capability is trained without them. \\
\addlinespace[3pt]
Objective
& KT-IPA \cite{cheng2025weaponization}
& T2\,+\,T3
& Map the residual-stream activations of harmful inputs onto random target vectors (T2), then train adversarially against sampled simulated attacks within a prospect-theoretic objective (T3). \\
\addlinespace[3pt]
Trajectory
& MLAC \cite{henderson2023self}
& T3
& Meta-train against $K$ inner harmful-adaptation steps of a meta-learned adversary, so harmful loss remains high averaged over the simulated trajectory. \\
& TAR \cite{tamirisa2024tamper}
& T3
& After an initial safeguard, meta-train against sampled full-parameter fine-tuning attacks of $K$ steps, optimizing an entropy or refusal-preference loss at points along each trajectory. \\
\midrule
\multicolumn{4}{@{}l}{\textit{\textbf{Strategy B: Self-destruction}}} \\
\midrule
---
& CTRAP \cite{yi2025ctrap}
& T3
& After one simulated harmful step, train the resulting model to predict a fixed error token on general dialogue data, tying harmful updates to capability collapse. \\
& SEAM \cite{wang2025self}
& T4
& Minimize the cosine similarity between harmful and benign gradients, so descending the harmful loss approximately ascends the benign loss. \\
& SDD \cite{chen2025sdd}
& T4
& Fine-tune on harmful prompts paired with unrelated high-quality answers, so later harmful fine-tuning is intended to damage general instruction following. \\
\bottomrule
\end{tabular}
\vspace{1em}
\caption{Extended mapping of fifteen MFT defenses with author-intent rationale. Templates T1--T4 are defined in \S\ref{sec:templates}. $^{\dagger}$LoX is the training-free analogue of T1: it does not optimize T1 directly, but its weight extrapolation targets the same flat-basin geometry.}
\label{tab:mft-defense-taxonomy-appendix}
\end{table}

\section{Compute resources}
\label{app:compute}
The cost-curve experiment comprises 96 attack trajectories of 3{,}375 optimizer steps each, for 324{,}000 steps in total. Its 2{,}016 evaluated checkpoints each generate responses to 1{,}000 harmful prompts and undergo the 600-question capability evaluation. These counts describe the reported trajectories; they exclude the earlier defense-training sweeps. We do not report a GPU-hour total for this experiment because the recorded training times do not include all generation, judging, and defense construction costs. Supervised completion-token counts measure training exposure and are not estimates of total compute.

\section{Experimental setup details}
\label{app:setup_details}

\noindent\textbf{Attack data and objective.} The source file contains 10{,}000 harmful BeaverTails training examples~\cite{ji2023beavertails}. The attack loader uses a 90/10 train/validation split, leaving 9{,}000 examples for training. The remaining 1{,}000 examples are not used for updates, checkpoint selection, or early stopping in these trajectories. The attack minimizes cross-entropy on the harmful completion, with prompt and padding tokens masked from the loss. It uses no benign-data term. All 96 trajectories use the same source file and split seed. The loader shuffles the training set once and cycles through the same order in later epochs.

\subsection{Attack hyperparameters and token accounting}
\label{app:hyperparams}

We use AdamW with peak learning rates $10^{-5}$, $3\times10^{-5}$, and $10^{-4}$, one per trajectory. The cosine schedule spans all 3{,}375 steps, with 337 warmup steps, weight decay zero, and gradient norm clipping at 1.0. The microbatch size is four, gradient accumulation is two, and the maximum sequence length is 512. LoRA uses rank 16, $\alpha=32$, dropout $0.05$, and the attention and MLP projections \texttt{q\_proj}, \texttt{k\_proj}, \texttt{v\_proj}, \texttt{o\_proj}, \texttt{gate\_proj}, \texttt{up\_proj}, and \texttt{down\_proj}. The attack seed is 42 throughout, and training continues for three full epochs without early stopping.

Checkpoints are recorded after steps 1, 2, 3, 5, 7, 10, 15, 22, 33, 48, 70, 104, 153, 225, 331, 488, 718, 1{,}057, 1{,}557, 2{,}292, and 3{,}375. At step $s$, the cumulative number of example presentations is $8s$, and the number of distinct training records encountered is $\min(8s,9{,}000)$. The first complete pass ends at step 1{,}125. We count supervised completion positions after tokenization and truncation, so the plotted token budgets differ across tokenizers. The vertical markers use the exact first-epoch token count from the undefended training log; a defended model's template can produce a slightly different count.

The first optimizer step uses a zero learning rate because warmup starts at zero. Its model weights therefore match the pre-attack weights, although the forward/backward pass has consumed tokens. We retain that recorded token count on the logarithmic axis. Later checkpoints are prefixes of a schedule configured for the full run, not independently optimized attacks for each shorter budget.

\subsection{Defense construction and checkpoint provenance}
\label{app:defense-provenance}

The curves start from six defended checkpoints per backbone, all with defense seed 42. We use author-released defense implementations with compatibility changes, data-format adapters, and training wrappers. Booster and CTRAP use the last checkpoints saved before validation-based early stopping ended their training runs. The undefended reference is the original released chat model, attacked with the same configuration.

For SAM-unlearning (Unlearn-Smooth in the figures), we port the authors' negative preference optimization (NPO) forget loss and sharpness-aware minimization (SAM) update into a trainer that supports all four backbones. The forget target is the harmful response in paired BeaverTails data; the retain objective uses safe responses and benign instructions. This changes the task from the authors' WMDP-Bio knowledge unlearning to safety alignment. The defense uses LoRA on the query and value projections, with rank 8 and $\alpha=32$. These locally trained NPO+SAM checkpoints are an adaptation of the method, rather than the author-released WMDP checkpoint.

For Vaccine, VAA, SDD, and SAM-unlearning, the defense configurations were selected in earlier sweeps using a validation objective combining harmfulness and capability. The capability component uses the same validation subsets plotted here, so it is not an independent test of the selected defense. The comparison concerns these particular instantiations of each defense, including differences in their starting capability.

\subsection{Evaluation protocol}

\noindent\textbf{Continuous harmfulness.} We take a fixed category-stratified sample of 1{,}000 prompts from the 18{,}189 unsafe prompts in the BeaverTails test split. The same prompt identifiers are used at every checkpoint. Responses use the model's chat template and the system prompt selected for that model, with greedy decoding and a limit of 512 new tokens. The fine-tuned StrongREJECT judge~\cite{souly2024strongreject}, \texttt{qylu4156/strongreject-15k-v1}, scores the saved responses. It normalizes the logits for grades 1 through 5, maps the grades to $0,0.25,0.5,0.75,1$, and returns their probability-weighted mean. The reported harmfulness is the average of these continuous scores over the 1{,}000 prompts. Judge input responses are truncated to 512 judge-tokenizer tokens. We do not threshold the resulting scores.

\noindent\textbf{Capability proxy.} The validation subsets contain 200 MMLU questions, 100 TruthfulQA questions, 200 HellaSwag questions, and 100 ARC-Easy questions. We score the next-token logits for the answer letters and select the highest-scoring option, with zero in-context examples. For TruthfulQA, the choices and correctness labels come from \texttt{mc2\_targets}; selecting any positively labeled option counts as correct. The plotted proxy is the equally weighted mean of the four task accuracies, not accuracy pooled over all 600 questions. It is not the standard full-suite \texttt{lm-eval-harness} metric, and its average can hide changes on individual tasks.

\section{Undefended models across backbones}
\label{app:undefended-curves}

\begin{figure}[htp]
\centering
\includegraphics[width=\linewidth]{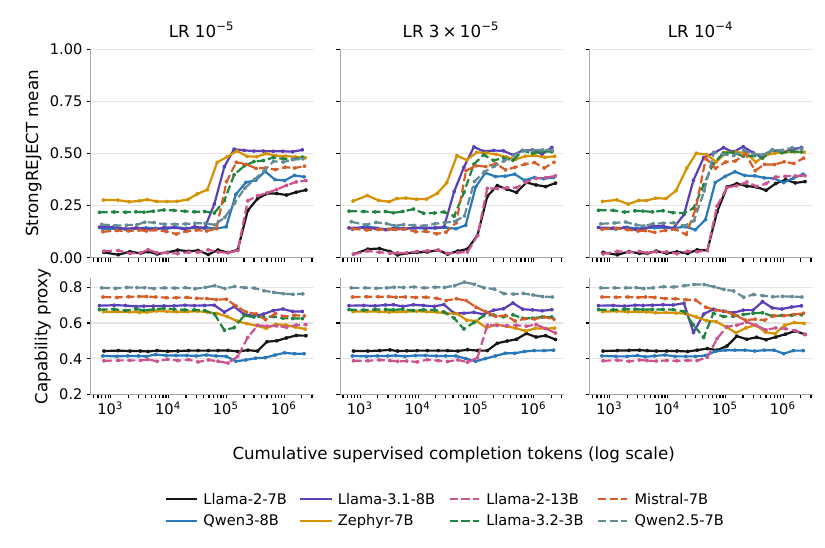}
\caption{Eight undefended models under continued MFT using the harmful-only objective. Colors identify models; each learning rate remains a separate column. Both axes use the same definitions as \autoref{fig:cost-llama2}. These are 24 trajectories, of which 12 also serve as references in the defended comparison.}
\label{fig:cost-undefended}
\end{figure}

\autoref{fig:cost-undefended} compares the undefended references from the main experiment with Llama-2-13B-chat, Llama-3.2-3B-Instruct, Mistral-7B-Instruct-v0.3, and Qwen2.5-7B-Instruct under the same data, schedule, and evaluation protocol. Each additional model contributes three trajectories. Absolute completion-token counts are tokenizer-dependent and are not a common FLOP or wall-clock budget across model sizes.

\end{document}